# Solar wind driven electrostatic instabilities with generalized (r,q) distribution function

**Warda Nasir[1], Zahida Ehsan[2*], M. N. S. Qureshi[3] and H. A. Shah[3]**

[1]Department of Physics, Forman Christian College, Lahore, 54000 Pakistan.

[2]Department of Physics, COMSATS Information of Technology, Lahore, 54000 Pakistan.

[3]Department of Physics, GC University, Lahore, 54000 Pakistan.

*For correspondence: ehsan.zahida@gmail.com

Using Boltzmann-Vlasov kinetic model with the flat-top non-Maxwellian distributed electron –ion, a velocity power law energetic tail, known as the generalized ($r$, $q$) distribution, a currentless electrostatic instability namely ion acoustic which is driven by a stream of solar wind plasma is studied here.

The instability threshold is affected and depends upon the spectral indices $r$ and $q$. It is found that the growth rate increases with the decrease of spectral index.

Moreover, such kinetic instability has also been discussed for a three species electron-ion-dust plasma using the generalized r, q distribution function. Such case is of interest when the solar wind is streaming through the cometary plasma in the presence of interstellar dust and excites electrostatic instabilities. In the limits of phase velocity of the waves larger and smaller than the thermal velocity of dust particles, the dispersion properties and growth rate of dust-acoustic mode are calculated analytically and has been plotted for different values of the spectral indices.

## I. INTRODUCTION

Ion acoustic wave can be excited and eventually become unstable when interpenetration of two plasmas takes place. Interpenetrating or so called *permeating* plasmas can be formed when for example one quasi-neutral plasma passes through another quasi-neutral relatively slower (or static plasma) [1,2]. Such satiation can arise quite frequently in laboratory and both space & astrophysical environment, for instance: there have been jets (about 24 events per second and are few thousand kilometers in diameter) which emerge in the transition region of the solar atmosphere. The jets were moving upwards with speeds of around 400 km/s. Similarly, upflows of plasma in numerous chromospheric spicules may reach any height in the corona between 5 x $10^3$ and 2 x $10^5$ km. Their diameters are typically greater than 1000 km, and they cover a few percent of the solar atmosphere in any moment.

Coronal rain though an opposite phenomenon, is also an example of interpenetrating plasmas. This is plasma moving down towards and through the lower layers of the solar atmosphere with almost a free-fall speed between 50 and 100 km/s [3, 4].

Moreover when solar wind passes through some cometry plasmas which contain dust particles in addition to usual electron proton plasma particulates and the region where two solar winds the fast and slower permeates also make up permeating plasmas [1].

In astrophysics, typical examples of such interpenetrating plasmas are clouds of novae and supernovae explosions spreading through the surrounding space or solar and stellar winds permeating through interstellar and interplanetary plasmas. In laboratory conditions, such interpenetrating plasmas with different parameters (e.g., pressure) can easily be realized [2, 3].

In general, it's believed that there is free energy available in such interpenetrating plasmas, which can drive various current-less electrostatic instabilities; some of them related to the ion acoustic wave instability. In a recent study of permeating plasma dynamics, a kinetic current-less instability was studied for two interpenetrating plasmas where the threshold velocity of the propagating plasma was found even below the IA speed of static plasma. In such (frequent) configurations in cometary magnetospheres, and interpenetrating space and astrophysical plasmas, this type of instability was found much more effective than in ordinary plasmas [3].

It is important to point out that for being currentless, it also eliminates the difficulties associated with the self induced magnetic field associated with the current due to background electrons [1]. The investigation of such kind of kinetic instability was carried out in the framework of thermodynamically stable plasma obeying the Maxwellian behavior.

However the observation of electrons in the space environment also shows that solar wind electrons and ions deviate considerably from the Maxwellian form, especially in the high energy range [5,6,7,8,9,10,11]. Vasyliunas (1968) introduced a phenomenological model known as the κ distribution, $f \sim (1 + v2/\kappa v^2_{th})^{-\kappa+1}$, in order to empirically fit the data [5]. Even though the κ distribution is a convenient tool [1,12], its theoretical justification is not on firm theoretical ground. Moreover, the choice of the parameter κ *(spectral indices)* is made arbitrarily.

Now it is possible but somewhat challenging due the complexities involved to construct the distribution function from the empirical data, which is found to deviate from the Maxwellian due to the presence of high energy tails and shoulders in the profile of the distribution functions.

In this respect, Summers et al. [13] and Nouman et al. [14] introduced the modified plasma dispersion functions, i.e., Z and Z r,q, which are analogous to the standard plasma dispersion function Z based on the Maxwellian. Since then there have been numerous studies where non-Maxwellian (*r, q*) distribution has been successfully used for the different treatments of plasmas, to mention just a few. Electrostatic modes and associated growth rates for usual electron ion and dusty plasma have been derived using r, q distribution [15] and later same group of authors studied electromagnetic (ordinary (O) and extraordinary (X)) modes and found that as the value of the spectral index *r* increases for a fixed value of *q*, the damping increases for the *O* mode but decreases for the *X* mode [16]. The problem of Weibel instability hs also been dealt with the generalized (r,q) distribution function [17].

The same has also been employed by Deba et al., [18] to study the electrostatic potential and energy loss of the projectiles showing that the presence of the high energy tail or of the shoulders modifies the results significantly. This distribution was also used to account for the solar wind heating due to the dissipation of Alfvén waves in space [19] and dusty plasmas [20].

In another attempt, a generalized dielectric constant for the electron Bernstein waves using non-Maxwellian distribution functions was derived in a collisionless, uniform magnetized plasma. Where the dispersion relations were dependent upon the spectral indices *(r, q)* and show that how the non-Maxwellian dispersion curves deviate from the Maxwellian depending upon the values of the spectral indices chosen. Authors also showed that the *(r, q)* dispersion relation is reduced to the kappa distribution for *r=0* and *q=κ+1*, which, in turn, is further reducible to the Maxwellian distribution for κ→∞ [21]. Later authors also observed that the propagation band for dust Bernstein waves is rather narrow as compared with that of the electron Bernstein waves. However, the band width increases for higher harmonics, for both kappa and (r,q) distributions [22].

The generalized r ,q distribution has produced very good fits for the magnetosheath electron data from the AMPTE satellite and the solar wind proton data from CLUSTER [23]. The (r, q) distribution is eminently suitable for modeling the electron distributions observed in the down-stream region of the terrestrial bow shock, but it can also be employed in a general situation in space plasmas. For instance, Zaheer and Yoon [2013] employed (r, q) distribution for modeling the solar wind electrons near 1 AU [24]. For the aforementioned investigations with the r, q distribution, we believe that choice of this destruction for the study of ion acoustic instability in permeating plasmas is justified and will lead to more accurate analysis. The distribution function we wuse in the proceedings sections I is the Generalized (r,q) distribution function in Cartesian coordinate system given as:

$$f_{j0} = \frac{3\,\Gamma[q](q-1)^{-\frac{3}{2(1+r)}}}{4\pi\,\theta_T^3\,\Gamma\left[q-\frac{3}{2(1+r)}\right]\Gamma\left[1+\frac{3}{2(1+r)}\right]} \left[1 + \frac{1}{(q-1)\theta_T^2}\left(v_x^2 + v_y^2 + (v_z - v_{j0})^2\right)^{(1+r)}\right]^{-q} \qquad (1)$$

Here, 'r', 'q' are the spectral indices representing the shoulder profile and high energy tail respectively having constraints q > 1 and q(1+r) > 5/2.This distribution function reduced to Maxwellian distribution function under the conditions "r = 0" & "q → ∞" and also to Lorentzian distribution when "r = 0", "q = κ+1". Where

$$\theta_T = \sqrt{\frac{T}{m}} \sqrt{\frac{3(q-1)^{-\frac{1}{(1+r)}} \Gamma\left(q - \frac{3}{2(1+r)}\right) \Gamma\left(\frac{3}{2(1+r)}\right)}{\Gamma\left(q - \frac{5}{2(1+r)}\right) \Gamma\left(\frac{5}{2(1+r)}\right)}}$$

is the thermal velocity. For the understating of the readers, it is worth mentioning that the parameters $r$ and $q$ in above equation generally represent the flat part and the high-energy tail of the distribution function, respectively.
Also for $r = 0$ and large values of $q$, the $(r,q)$ distribution exhibits the same behavior as that of usual kappa distribution. However if the value of $q$ is fixed and $r$ is increases, then the contribution of high-energy particles reduces and the shoulders become more prominent in the distribution function.

In addition to above we also plan to investigate a current less kinetic instability of dust-acoustic wave which is driven when solar wind permeates a cometry (dusty) plasma a situation also relevant when cometary plasma trail passing through the interplanetary space where dust is found ubiquitously.
The manuscript is organized as follows. In Sec. II, we present the basic formulation for the dynamics of ion acoustic wave via the dielectric function and obtain the dispersion equation and the instability condition. Whereas in section III, analysis of dust acoustic kinetic instability is given. Section IV is devoted to the quantitative analysis and conclusions on the ion (dust)-acoustic wave instability driven by solar wind plasmas with the generalized (r, q) distribution function.

## II. DISPERSION AND GROWTH OF IA WAVES With (r, q) DISTRIBUTION.

This is a kinetic derivation within which the plasma distribution function for the species j is used in the form
To start our calculations we take the kinetic Vlasov-Boltzmann equation for linearized perturbed state

$$\frac{\partial f_{j1}}{\partial t} + \vec{v} \cdot \frac{\partial f_{j1}}{\partial \vec{r}} + \frac{q_j}{m_j} \vec{E_1} \frac{\partial f_{j0}}{\partial \vec{v}} = 0 \qquad (2)$$

Also apply plane wave solution to above equation, we may derive

$$f_{j1} = -\frac{q_j}{m_j} \frac{k \varphi_1}{(\omega - k.v)} \frac{\partial f_{j0}}{\partial v} \qquad (3)$$

Using the above distribution function given in (1) and substitute it in (3) and defining the perturbed density as

$$n_{j1} = \int_{-\infty}^{\infty} f_{j1} \, d^3\vec{v} \qquad (4)$$

Now by using $f_{j1}$ here and doing simple arithmetic the general dispersion relation for ion acoustic wave may be given as

$$\frac{n_{j1}}{n_{j0}} = -\frac{e \phi_1}{k T_j C_1} \left[ B - A \int_{-\infty}^{\infty} \frac{\alpha_j}{(\alpha_j - \xi)} \left\{ 1 + \frac{1}{q-1} (\xi)^{2(r+1)} \right\} d\xi \right] \qquad (5)$$

Here A, B and $C_1$ are the constants that are equal to

$$A = \frac{(1+r)(q-1)^{-\frac{3}{2(1+r)}}\Gamma(q)}{2\Gamma\left(q-\frac{3}{2(1+r)}\right)\Gamma\left(\frac{3}{2(1+r)}\right)} \tag{6}$$

$$B = \frac{(q-1)^{-\frac{1}{1+r}}\Gamma\left(q-\frac{1}{2(1+r)}\right)\Gamma\left(\frac{1}{2(1+r)}\right)}{2\Gamma\left(q-\frac{3}{2(1+r)}\right)\Gamma\left(\frac{3}{2(1+r)}\right)} \tag{7}$$

$$C_{22} = \frac{3(q-1)^{-\frac{1}{1+r}}\Gamma\left(q-\frac{3}{2(1+r)}\right)\Gamma\left(\frac{3}{2(1+r)}\right)}{2\Gamma\left(q-\frac{5}{2(1+r)}\right)\Gamma\left(\frac{5}{2(1+r)}\right)} \tag{8}$$

and

$$Z_{r,q}(a_j) = A\int_{-\infty}^{\infty} \frac{a_j}{(a_j-\xi)}\left\{1+\frac{1}{q-1}(\xi)^{2(r+1)}\right\}d\xi \tag{9}$$

is the plasma dispersion function for *j* species (both electrons and ions). Here we use $a_j$ as the argument of this function for streaming $a_j = \frac{\omega}{kv_{Tj}}$ whereas $b_j$ as an argument for non-streaming $b_j = \frac{(\omega-kv_0)}{kv_{Tj}}$ species. By setting the following perturbed quasi-neutrality condition for deriving the cometary 'c' and wind 'w' plasma dispersion relation

$$n_{wi1} + n_{ci1} = n_{we1} + n_{ce1} \tag{10}$$

So while driving the dispersion relation for ion-acoustic wave for this case some expansions are, for both electrons populations we have $|a_e| \equiv \frac{|\omega|}{kv_{Tce}} \ll 1$ and $|b_e| \equiv \frac{|\omega-kv_0|}{kv_{Twe}} \ll 1$ and for the case in cometary ion the situation is like $|a_i| \equiv \frac{|\omega|}{kv_{Tci}} \ll 1$ whereas for the case of wind ions two situations may occur in which its one term $\left(\frac{\omega}{kv_{Twi}}\right) \ll 1$ and $\frac{v_0}{v_{Twi}} \gg 1$. For these we may use the asymptotic and power series as;

For $a_{ce} \ll 1$ (slow electrons)

$$Z(a_{ce}) = -\frac{3i\pi(q-1)^{-\frac{3}{2(1+r)}}\Gamma(q)}{4\Gamma\left(q-\frac{3}{2(1+r)}\right)\Gamma\left(1+\frac{3}{2(1+r)}\right)}a_{ce} \tag{11}$$

For $a_{ci} \ll 1$ (slow ions)

$$Z(a_{ci}) = -\frac{3i\pi(q-1)^{-\frac{3}{2(1+r)}}\Gamma(q)}{4\Gamma\left(q-\frac{3}{2(1+r)}\right)\Gamma\left(1+\frac{3}{2(1+r)}\right)}a_{ci} \tag{12}$$

For $b_{we} \ll 1$ (Wind or flowing electrons)

$$Z(b_{we}) = -\frac{3 i \pi (q-1)^{-\frac{3}{2(1+r)}} \Gamma(q)}{4 \Gamma\left(q-\frac{3}{2(1+r)}\right)\Gamma\left(1+\frac{3}{2(1+r)}\right)} a_{we} \tag{13}$$

For $b_{wi} \ll 1$ (Wind ions)

$$Z(b_{wi}) = -\frac{3 i \pi (q-1)^{-\frac{3}{2(1+r)}} \Gamma(q)}{4 \Gamma\left(q-\frac{3}{2(1+r)}\right)\Gamma\left(1+\frac{3}{2(1+r)}\right)} b_{wi} \left[1+\frac{b_{wi}^{2(1+r)}}{q-1}\right]^{-q} + \frac{(q-1)^{-\frac{3}{2(1+r)}}}{\Gamma\left(q-\frac{3}{2(1+r)}\right)\Gamma\left(\frac{3}{2(1+r)}\right)}\left\{(q-1)^{\frac{1}{2(1+r)}}\Gamma\left(q-\frac{1}{2(1+r)}\right)\Gamma\left(\frac{1}{2(1+r)}\right) + \frac{(q-1)^{\frac{3}{2(1+r)}}\Gamma\left(q-\frac{3}{2(1+r)}\right)\Gamma\left(\frac{3}{2(1+r)}\right)}{b_{wi}^2} + \cdots\right\}$$

(14)

Hence by planting all these expansion in eq. (5) the dispersion relation takes the form:

$$-Z_{wi}^2 \frac{n_{wi0}}{T_{wi}}\left\{\frac{k^2 \psi_{Twi}^2}{(\omega-kv_0)^2} - i\pi A \frac{(\omega-kv_0)}{k\psi_{Twi}}\left[1+\frac{\beta_{wi}^{2(1+r)}}{q-1}\right]^{-q}\right\} + Z_{ci}^2 \frac{n_{ci0}}{T_{ci}}\left(B+i\pi A \frac{\omega}{k\psi_{Tci}}\right)$$
$$+\frac{n_{we0}}{T_{we}}\left(B+i\pi A \frac{(\omega-kv_0)}{k\psi_{Twe}}\right) + \frac{n_{ce0}}{T_{ce}}\left(B+i\pi A \frac{\omega}{k\psi_{Tce}}\right) \equiv Re[\Delta(k,\omega)] + iIm[\Delta(k,\omega)] = 0$$

(15)

It seems to be like as the ions being gigantic the ion influence in eq. (15) will be deserted and is totally omitted from the wind ions of imaginary part. Also the term due to cometary electrons is neglected as it is very less than the cometary ions. Therefore, by sorting out the above equation in real and imaginary parts the relation for real frequency can be gained as:

$$\omega_r^2 = k^2 Z_{wi}^2 \frac{n_{wi0}}{n_{ce0}} \frac{k_B T_{ce}}{m_{wi}} \frac{1}{B} \frac{1}{1+\frac{n_{we0} T_{ce}}{n_{ce0} T_{we}}+Z_{ci}^2 \frac{n_{ci0} T_{ce}}{n_{ce0} T_{ci}}} \tag{16}$$

Whereas; the normalized value of real frequency is of the form

$$\frac{\omega_r^2}{\omega_{pi}^2} = k^2 \lambda_{Dwi}^2 Z_{wi} \frac{n_{wi0}}{n_{ce0}} \frac{T_{ce}}{T_{wi} c_{22}} \frac{1}{B} \frac{1}{1+\frac{n_{we0} T_{ce}}{n_{ce0} T_{we}}+Z_{ci}^2 \frac{n_{ci0} T_{ce}}{n_{ce0} T_{ci}}} \tag{17}$$

Whereas, the relation for growth rate can be attained by the following relation [Baumjohann et al., 1997]:

$$\gamma = -\frac{Im(\Delta)}{\frac{\partial Re(\Delta)}{\partial \omega_r}}$$

So

$$\gamma = \frac{\pi}{2} \frac{A}{z_{wi}^2} \frac{n_{we0}}{n_{wi0}} \frac{m_{wi} m_{we}^{\frac{1}{2}}}{(C_{22} k_B T_{we})^{\frac{3}{2}}} \frac{\omega_r^3}{k^2} \left[ v_0 - \frac{\omega_r}{k} \left\{ 1 + Z_{ci}^2 \frac{n_{ci0}}{n_{we0}} \left(\frac{T_{we}}{T_{ci}}\right)^{\frac{3}{2}} \left(\frac{m_{ci}}{m_e}\right)^{\frac{1}{2}} \right\} \right] \qquad (18)$$

The normalized growth rate equation can be written as;

$$\frac{\gamma}{\omega_{pi}} = \frac{\pi}{2} \frac{A}{k^2 \lambda_{Dwi}^2} \frac{n_{we0}}{n_{wi0} Z_{wi}} \frac{T_{wi}}{T_{we}} \frac{\omega_r^3}{\omega_{pi}^3} \left[ v_0 - \frac{\omega_r/\omega_{pwi}}{k \lambda_{Dwi}} \sqrt{\frac{Z_{wi} T_{wi} m_{we}}{T_{we} m_{wi}}} \left\{ 1 + Z_{ci}^2 \frac{n_{ci0}}{n_{we0}} \left(\frac{T_{we}}{T_{ci}}\right)^{\frac{3}{2}} \left(\frac{m_{ci}}{m_e}\right)^{\frac{1}{2}} \right\} \right] \qquad (19)$$

With the threshold instability condition,

$$v_0 > \frac{\omega_r}{k} \left( 1 + Z_{ci}^2 \frac{n_{ci0}}{n_{ce0}} \left(\frac{T_{we}}{T_{ci}}\right)^{\frac{3}{2}} \left(\frac{m_{ci}}{m_e}\right)^{\frac{1}{2}} \right) \qquad (20)$$

## III. DISPERSION AND GROWTH OF DUST ACOUSTIC WAVES IN NON-MAXWELLIAN PLASMAS

Now, we consider our permeating plasma comprises a flowing (e.g., solar wind) and relatively stationary or target plasma (e.g., cometary dusty plasma). Let the background number densities are n0xj (solar wind plasma), n0cj (cometary plasma), and n0d (static dust grains), where the index j¼e and i stand for electrons and ion species.

By starting with Kinetic Vlasov-Boltzmann Equation in perturbed linearized distribution function

$$\frac{\partial f_{j1}}{\partial t} + \vec{v} \cdot \frac{\partial f_{j1}}{\partial \vec{r}} + \frac{q_j}{m_j} \vec{E_1} \frac{\partial f_{j0}}{\partial \vec{v}} = 0 \qquad (21)$$

and substituting the Generalized ($r,q$) distribution function is of the form

$$f_{j0} = \frac{3 \Gamma[q](q-1)^{-\frac{3}{2(1+r)}}}{4 \pi \theta_T^3 \Gamma\left[q - \frac{3}{2(1+r)}\right] \Gamma\left[1 + \frac{3}{2(1+r)}\right]} \left[ 1 + \frac{1}{(q-1)\theta_T^2} \left(v_x^2 + v_y^2 + (v_z - v_{j0})^2\right)^{(1+r)} \right]^{-q} \qquad (22)$$

r and q are the spectral indices where r represents the shoulder profile and q is for high energy tail. Now applying plane wave solution and solving for $f_{ji}$ and introducing the density perturbation equation

$$n_{j1} = \int_{-\infty}^{\infty} f_{j1} \, d^3\vec{v} \qquad (23)$$

one may derive the dispersion relation for ion acoustic wave as:

$$\frac{n_{j1}}{n_{j0}} = -\frac{e \phi_1}{k T_j C_{22}} \left[ B - A \int_{-\infty}^{\infty} \frac{\alpha_j}{(\alpha_j - \xi)} \left\{ 1 + \frac{1}{q-1} (\xi)^{2(r+1)} \right\} \right] d\xi \qquad (24)$$

Here A, B and $C_1$ are the constants that are equal to

$$A = \frac{(1+r)(q-1)^{-\frac{3}{2(1+r)}}\Gamma(q)}{2\Gamma\left(q-\frac{3}{2(1+r)}\right)\Gamma\left(\frac{3}{2(1+r)}\right)}$$

$$B = \frac{(q-1)^{-\frac{1}{1+r}}\Gamma\left(q-\frac{1}{2(1+r)}\right)\Gamma\left(\frac{1}{2(1+r)}\right)}{2\Gamma\left(q-\frac{3}{2(1+r)}\right)\Gamma\left(\frac{3}{2(1+r)}\right)}$$

$$C_{22} = \frac{3(q-1)^{-\frac{1}{1+r}}\Gamma\left(q-\frac{3}{2(1+r)}\right)\Gamma\left(\frac{3}{2(1+r)}\right)}{2\Gamma\left(q-\frac{5}{2(1+r)}\right)\Gamma\left(\frac{5}{2(1+r)}\right)}$$

and

$$Z_{r,q}(\alpha_j) = A\int_{-\infty}^{\infty}\frac{\alpha_j}{(\alpha_j-\xi)}\left\{1+\frac{1}{q-1}(\xi)^{2(r+1)}\right\}d\xi$$

is the plasma dispersion function for $j$ species (both electrons and ions). Here we use $\alpha_j$ as the argument of this function for streaming $\alpha_j = \frac{\omega}{kv_{Tj}}$ whereas $\beta_j$ as an argument for non-streaming $\beta_j = \frac{(\omega-kv_0)}{kv_{Tj}}$ species. By setting the following perturbed quasi-neutrality condition for deriving the cometary and wind plasma dispersion relation

$$n_{wi1} + n_{ci1} = n_{we1} + n_{ce1} + z_d n_{d1}$$

So while driving the dispersion relation for ion-acoustic wave for this case some expansions are used under the conditions $|\alpha_d| \gg 1$ and $|Re\,(\alpha)_d| \gg |Im\,(\alpha)_d|$, for both electrons populations we have $|\alpha_e| \equiv \frac{|\omega|}{kv_{Tce}} \ll 1$ and $|\beta_e| \equiv \frac{|\omega-kv_0|}{kv_{Twe}} \ll 1$ and for the case in cometary ion the situation is like $|\alpha_i| \equiv \frac{|\omega|}{kv_{Tci}} \ll 1$ whereas for the case of wind ions two situations may occur in which its one term $\left(\frac{\omega}{kv_{Twi}}\right) \ll 1$ and $\frac{v_0}{v_{Twi}} \gg 1$.

For $\alpha_{ce} \ll 1$ (Cometary electrons)

$$Z(\alpha_{ce}) = -\frac{3i\pi(q-1)^{-\frac{3}{2(1+r)}}\Gamma(q)}{4\Gamma\left(q-\frac{3}{2(1+r)}\right)\Gamma\left(1+\frac{3}{2(1+r)}\right)}\alpha_{ce} \tag{25}$$

For $\alpha_{ci} \ll 1$ (Cometary ions)

$$Z(\alpha_{ci}) = -\frac{3i\pi(q-1)^{-\frac{3}{2(1+r)}}\Gamma(q)}{4\Gamma\left(q-\frac{3}{2(1+r)}\right)\Gamma\left(1+\frac{3}{2(1+r)}\right)}\alpha_{ci} \tag{26}$$

For $\beta_{we} \ll 1$ (Wind electrons)

$$Z(\beta_{we}) = -\frac{3i\pi\,(q-1)^{-\frac{3}{2(1+r)}}\Gamma(q)}{4\Gamma\left(q-\frac{3}{2(1+r)}\right)\Gamma\left(1+\frac{3}{2(1+r)}\right)}\beta_{we} \tag{27}$$

For $\beta_{wi} \ll 1$ (Wind ions)

$$Z(\beta_{wi}) = -\frac{3i\pi\,(q-1)^{-\frac{3}{2(1+r)}}\Gamma(q)}{4\Gamma\left(q-\frac{3}{2(1+r)}\right)\Gamma\left(1+\frac{3}{2(1+r)}\right)}\beta_{wi}\left[1+\frac{\beta_{wi}^{2(1+r)}}{q-1}\right]^{-q} + \frac{(q-1)^{-\frac{3}{2(1+r)}}}{\Gamma\left(q-\frac{3}{2(1+r)}\right)\Gamma\left(\frac{3}{2(1+r)}\right)}\left\{(q-1)^{\frac{1}{2(1+r)}}\Gamma\left(q-\frac{1}{2(1+r)}\right)\Gamma\left(\frac{1}{2(1+r)}\right) + \frac{(q-1)^{\frac{3}{2(1+r)}}\Gamma\left(q-\frac{3}{2(1+r)}\right)\Gamma\left(\frac{3}{2(1+r)}\right)}{\beta_{wi}^2} + \cdots\right\}$$

(28)

For $\alpha_d \ll 1$ (Dust particles)

$$Z(\alpha_d) = -\frac{3i\pi\,(q-1)^{-\frac{3}{2(1+r)}}\Gamma(q)}{4\Gamma\left(q-\frac{3}{2(1+r)}\right)\Gamma\left(1+\frac{3}{2(1+r)}\right)}\alpha_d\left[1+\frac{\alpha_d^{2(1+r)}}{q-1}\right]^{-q} + \frac{(q-1)^{-\frac{3}{2(1+r)}}}{\Gamma\left(q-\frac{3}{2(1+r)}\right)\Gamma\left(\frac{3}{2(1+r)}\right)}\left\{(q-1)^{\frac{1}{2(1+r)}}\Gamma\left(q-\frac{1}{2(1+r)}\right)\Gamma\left(\frac{1}{2(1+r)}\right) + \frac{(q-1)^{\frac{3}{2(1+r)}}\Gamma\left(q-\frac{3}{2(1+r)}\right)\Gamma\left(\frac{3}{2(1+r)}\right)}{\alpha_d^2} + \cdots\right\}$$

(29)

Hence by planting all these expansion in eq. (24) the dispersion relation takes the form:

$$-Z_{wi}^2\frac{n_{wi0}}{T_{wi}}\left\{\frac{k^2\psi_{Twi}^2}{(\omega-kv_0)^2} - i\pi A\frac{(\omega-kv_0)}{k\psi_{Twi}}\left[1+\frac{\beta_{wi}^{2(1+r)}}{q-1}\right]^{-q}\right\} + Z_{ci}^2\frac{n_{ci0}}{T_{ci}}\left(B + i\pi A\frac{\omega}{k\psi_{Tci}}\right) + \frac{n_{we0}}{T_{we}}\left(B + i\pi A\frac{(\omega-kv_0)}{k\psi_{Twe}}\right) + \frac{n_{ce0}}{T_{ce}}\left(B + i\pi A\frac{\omega}{k\psi_{Tce}}\right) - Z_d^2\frac{n_{d0}}{T_d}\left\{\frac{k^2\psi_{Td}^2}{\omega^2} - i\pi A\frac{\omega}{k\psi_{Td}}\left[1+\frac{\alpha_d^{2(1+r)}}{q-1}\right]^{-q}\right\} \equiv Re[\Delta(k,\omega)] + iIm[\Delta(k,\omega)] = 0$$

(30)

It seems to be like as the ions being gigantic the ion influence in eq. (30) will be deserted and is totally omitted from the wind ions of imaginary part. Also the term due to cometary electrons is neglected as it is very less than the cometary ions. Therefore, by sorting out the above equation in real and imaginary parts the relation for real frequency can be gained as:

$$\omega_r^2 = k^2 Z_d^2\frac{n_{d0}}{n_{ce0}}\frac{kT_{ce}}{m_d}\frac{C_{22}}{B}\frac{1}{1+\frac{n_{we0}T_{ce}}{n_{ce0}T_{we}}+Z_{ci}^2\frac{n_{ci0}T_{ce}}{n_{ce0}T_{ci}}} \tag{31}$$

Normalized real frequency equation is

$$\frac{\omega_r^2}{\omega_{pd}^2} = k^2\lambda_{Dd}^2 Z_d\frac{n_{d0}}{n_{ce0}}\frac{T_{ce}}{T_d c_{22}}\frac{1}{B}\frac{1}{1+\frac{n_{we0}T_{ce}}{n_{ce0}T_{we}}+Z_{ci}^2\frac{n_{ci0}T_{ce}}{n_{ce0}T_{ci}}} \tag{32}$$

Whereas, the relation for growth rate can be attained by the following relation [Baumjohann et al., 1997]:

$$\gamma = -\frac{Im(\Delta)}{\frac{\partial Re(\Delta)}{\partial \omega_r}}$$

So

$$\gamma = \frac{\pi}{2} \frac{A}{z_d^2} \frac{n_{we0}}{n_{d0}} \frac{m_d m_{we}^{\frac{1}{2}}}{(C_{22} k T_{we})^{\frac{3}{2}}} \frac{\omega_r^3}{k^2} \left[ v_0 - \frac{\omega_r}{k} \left\{ 1 + Z_{ci}^2 \frac{n_{ci0}}{n_{we0}} \left(\frac{T_{we}}{T_{ci}}\right)^{\frac{3}{2}} \left(\frac{m_{ci}}{m_{we}}\right)^{\frac{1}{2}} \right\} \right] \quad (33)$$

The normalized growth rate equation is

$$\frac{\gamma}{\omega_{pd}} = \frac{\pi}{2} \frac{A}{k^2 \lambda_{Dd}^2} \frac{n_{we0}}{n_{d0} Z_d} \frac{T_d}{T_{we}} \frac{\omega_r^3}{\omega_{pd}^3} \left[ \frac{v_0}{v_{thwe}} - \frac{\omega_r/\omega_{pd}}{k \lambda_{Dd}} \sqrt{\frac{Z_d T_d m_{we}}{T_{we} m_d}} \left\{ 1 + Z_{ci}^2 \frac{n_{ci0}}{n_{we0}} \left(\frac{T_{we}}{T_{ci}}\right)^{\frac{3}{2}} \left(\frac{m_{ci}}{m_{we}}\right)^{\frac{1}{2}} \right\} \right] \quad (34)$$

With the threshold instability condition,

$$v_0 > \frac{\omega_r'}{k} \left( 1 + \frac{1}{Z_{ci}^2 \frac{n_{ci0}}{n_{we0}} \left(\frac{T_{we}}{T_{ci}}\right)^{\frac{3}{2}} \left(\frac{m_{ci}}{m_{we}}\right)^{\frac{1}{2}}} \right) \quad (35)$$

where;

$$\omega_r' = \sqrt{k^2 Z_d^2 \frac{n_{d0}}{n_{ce0}} \frac{k T_{ce}}{m_d} \frac{C_{22}}{B} \frac{1}{1 + \frac{n_{we0}}{n_{ce0}} \frac{T_{ce}}{T_{we}} + Z_{ci}^2 \frac{n_{ci0}}{n_{ce0}} \frac{T_{ce}}{T_{ci}}} \left( Z_{ci}^2 \frac{n_{ci0}}{n_{we0}} \left(\frac{T_{we}}{T_{ci}}\right)^{\frac{3}{2}} \left(\frac{m_{ci}}{m_{we}}\right)^{\frac{1}{2}} \right)}$$

## IV. QUANTITATIVE ANALYSIS AND CONCLUSIONS

Using the kinetic theory, we have derived the general dispersion relation for IA wave instability given in Eq. (16) in case of permeating plasmas based on $r$, $q$ distribution which is when solar wind plasma permeates through the relatively hotter and static space plasma. Later the associated growth rate of the IA current less instability was obtained (See (18)) which is dependent on spectral indices i.e., $r$, and $q$. Finally, threshold condition for the IA instability was also derived which clearly indicates that the condition is changed for the non-Maxwellian plasma.

For the case of a generalized ($r$, $q$) distribution function we have carried out a similar analysis for the dust acoustic current less kinetic instability which is triggered when for example solar wind stream passes through the cometry plasma omnipresent in the space. For this case we obtain dispersion relation given by (31) and instability growth rate (33) and threshold condition for the instability set up presented in (35) which are dependent upon the spectral indices $r$ and $q$.
To validate the results found in this manuscript, we make use of the typical parameters relevant to the observed situations. Particularly, we analyze the impact of the spectral index and streaming velocity on the instability growth of the non- Maxwellian permeating plasma mode without and with the presence of additional dust species.

We observe that when the growth rates of IA and DA instability are plotted against normalized propagation vector for different values of $r$ and $q$, the waves damping increases as compared to Maxwellian plasma as the value of the spectral index increases changes.

Fig. (1) represents the growth rate which is plotted against the normalized wave number such that for the different values of r, q, i.e., $(r, q) = (2.3, 1.2)$ shown in green line, $r = 3.0$ and $q = 1.7$ the red line, black line shows $(r, q) = (1.5, 2)$ $r = 2$, $q = 1.9$ are by the blue line for fixed value of streaming velocity: $v_0 = 8 \times 10^7$ cm/s. Whereas for $r = 0$ and very large $q$ the brown line indicates the case of Maxwellian plasma.

From here we may justify our results that by using such kind of distribution functions the growth rate decreases in comparison with Maxwellian distribution function. Hence the frequency is less than the frequency of Maxwellian growth rate.

On the other hand Fig. (2) Shows the growth rates of dust acoustic instability at various spectral indices values and the brown line indicating the usual Maxwellian case. We note that in this case due to the r, q distribution the frequency becomes larger than the Maxwellian case.

Conclusively we see a change in the dispersion relations and associated growth rates of the electrostatic currentless instabilities driven by the stream of solar wind so we believe that the use of generalized (r,q) distribution function is a better choice than maxwellian and kappa type of distributions [1,3].

## References:

[1] K. Arshad, Z. Ehsan, S. A. Khan, and S. Mahmood, Phys. Plasmas 21, 023704 (2014).

[2] J. Gong, Z. Liu, and J. Du Phys. Plasmas 19, 083706 (2012).

[3] J. Vranjes, S. Poedts, and Z. Ehsan, Phys. Plasmas 16, 074501 (2009).

[3] A. O. Benz, Plasma Astrophysics: Kinetic Processes in Solar and Stellar Coronae (Kluwer, Dordrecht, 2002), p. 106, 279.

[4] G. E. Brueckner and J. D. F. Bartoe, Astrophys. J. 272, 329 (1983).

[5] V. M. Vasyliunas, *J. Geophys. Res.*, 73, 2839 (1968).

[6] W. C. Feldman, J. R. Asbridge, S. J. Bame, M. D. Montgomery, & S. P. Gary, *J. Geophys. Res.*, 80, 4181 (1975).

[7] J. T. Gosling, J. R. Asbridge, S. J. Bame, W. C. Feldman, R. D. Zwickl, G. Paschmann, N. Sckopke, and R. J. Hynd *J. Geophys. Res.*, 86, 547 (1981).

[8] R. P. Lin, W. K. Levedahl, W. Lotko, D. A. Gurnett and F. L. Scarf, Astrophys. J, 308, 954 (1986).

[9] T. P. Armstrong, M. T. Paonessa E. V. II. Bell, and S. M. Krimigis, *J. Geophys. Res.*, 88, 8893 (1983).

[10] S. R. Cranmer, Astrophys. J 508, 925 939, (1998).

[11] R. E. Ergun, D. Larson, R. P. Lin, J. P. McFadden, C. W. Carlson, K. A. Anderson, L. Muschietti, M. McCarthy, G. K. Parks, H. Reme, J. M. Bosqued, C. D'Uston, T. R. Sanderson, K. P. Wenzel, Michael Kaiser, R. P. Lepping, S. D. Bale, Paul Kellogg, and J.-L. Bougeret Astrophys. J., 503, 435 (1998).

[12] O.R. Rufai, A.S. Bains, and Z. Ehsan. Astrophys Space Sci357:102 (2015).

[13] D. Summers and R. M. Thorne, Phys. Fluids B 3(8), 1835 (1991).

[14] M. N. S. Qureshi, H. A. Shah, G. Murtaza, F. Mahmood, and S. J. Schwartz, Phys. Plasmas 11, 3819 (2004).

[15] S. Zaheer, G. Murtaza, and H. A. Shah, Phys. Plasmas 11, 2246 (2004).

[16] S. Zaheer, G. Murtaza, H. A. Shah. Physics of Plasmas 13:6, 062109 (2006).

[17] S. Zaheer, G. Murtaza  Phys. Plasmas 14, 022108 (2007).

[18] F. Deeba, Z. Ahmad, and G. Murtaza, Phys. Plasmas 13, 082108 (2006).

[19] K. Zubia, H. A. Shah, M. N. S. Qureshi, and G. Murtaza, Sol. Phys. 236, 167 (2006).

[20] K. Zubia, H. A. Shah, P. H. Yoon Phys. Plasmas 22 8, 082902 (2015).

[21] F. Deeba, Zahoor Ahmad, and G. Murtaza , Phys. Plasmas **17**, 102114 (2010).

[22] F. Deeba, Zahoor Ahmad, and G. Murtaza, Phys. Plasmas **18**, 072104 (2011).

[23] M. N. S. Qureshi, W. Nasir, W. Masood, P. H. Yoon, H. A. Shah, and S. J. Schwartz, *J. Geophys. Res. Space Physics*, *119*, (2014): doi:10.1002/2014JA020476.

[24] S. Zaheer and P. H. Yoon Astrophys.  J., 775, 108 (2013) doi:10.1088/0004-637X/775/2/108.

**Acknowledgments:**
One of us (Z.E.) acknowledges Professor Stefaan Poedts and and Dr. Jovo Vranjes for the fruitful discussions.

**Figure captions:**

1. Fig (1) pictures the growth rate of ion acoustic instability for generalized (*r, q*) distribution function for for different *r* and *q* at fixed solar wind velocity of  $v_0$=8x10$^7$ cm/s.
2. Fig. (2): Growth rate of DA instability at different values of *r* and *q* at $v_0$=8x10$^7$ cm/s.

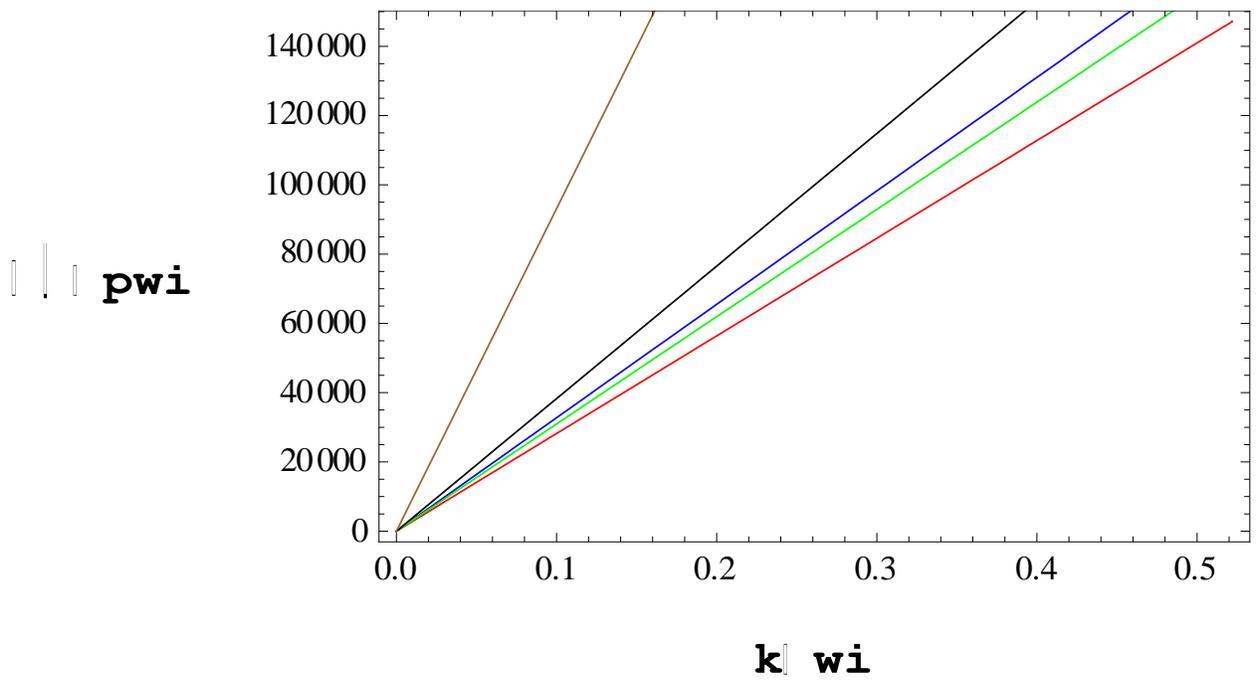

Figure (1)

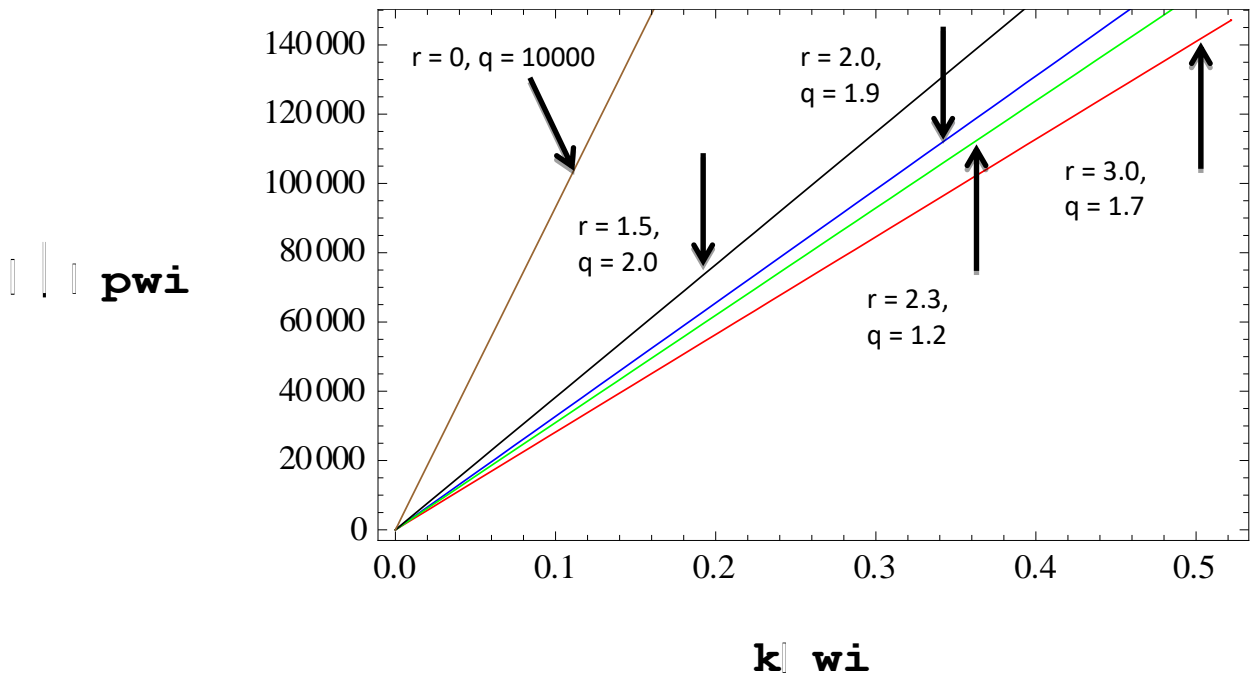

**Figure 1**: Growth rate at different r and q values particles

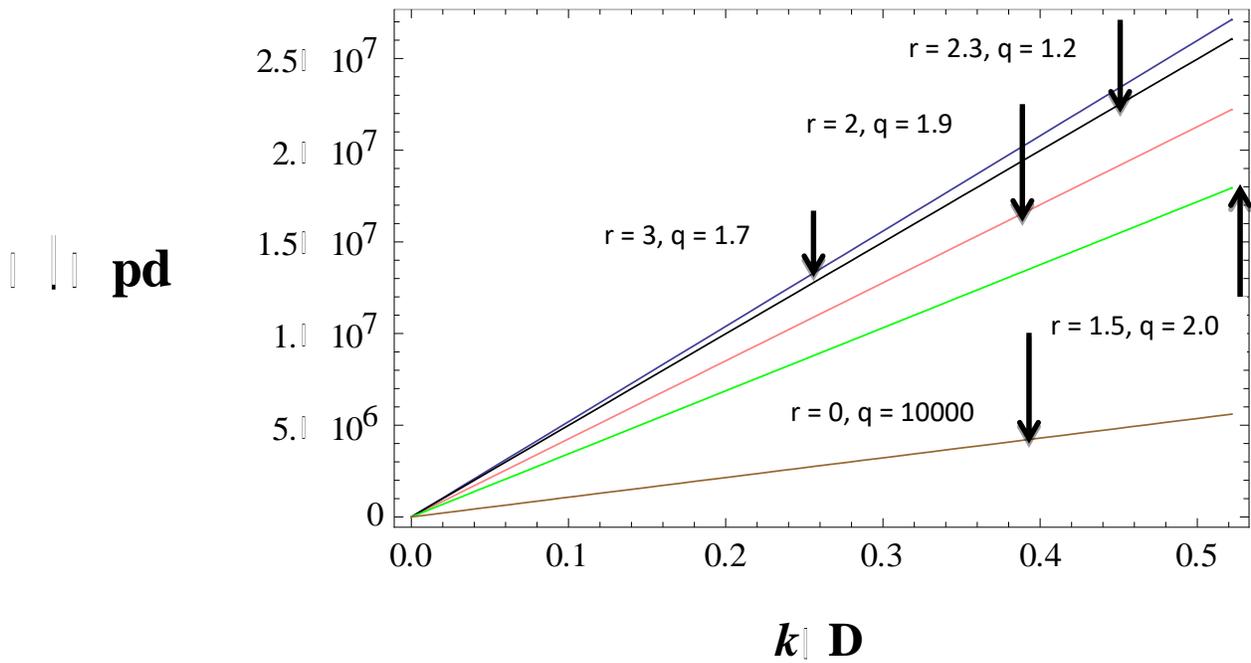

Figure (2)